\documentclass[journal=nalefd,manuscript=letter]{achemso}
\usepackage{graphicx}
\usepackage{array}
\usepackage{upgreek}
\usepackage{fixltx2e}
\usepackage{bibunits}

\title{Electric-field control of interfering transport pathways in a single-molecule anthraquinone transistor}

\author{Max Koole}
\affiliation[Delft University of Technology]
{Kavli Institute of Nanoscience, Delft University of Technology, Lorentzweg 1, 2628 CJ, Delft, The Netherlands}
\author{Jos M. Thijssen}
\affiliation[Delft University of Technology]
{Kavli Institute of Nanoscience, Delft University of Technology, Lorentzweg 1, 2628 CJ, Delft, The Netherlands}
\author{Hennie Valkenier}
\affiliation[University of Groningen]
{Stratingh Institute for Chemistry and Zernike Institute for Advanced Materials, University of Groningen, Nijenborgh 4, 9747 AG, Groningen, The Netherlands}
\alsoaffiliation[University of Bristol]
{Current address: School of Chemistry, University of Bristol, Cantock's Close, Bristol BS8 1TS, United Kingdom}
\author{Jan C. Hummelen}
\affiliation[University of Groningen]
{Stratingh Institute for Chemistry and Zernike Institute for Advanced Materials, University of Groningen, Nijenborgh 4, 9747 AG, Groningen, The Netherlands}
\author{Herre S.J. van der Zant}
\affiliation[Delft University of Technology]
{Kavli Institute of Nanoscience, Delft University of Technology, Lorentzweg 1, 2628 CJ, Delft, The Netherlands}
\email{H.S.J.vanderZant@tudelft.nl}


\begin{document}

\begin{abstract}
	It is understood that molecular conjugation plays an important role in charge transport through single-molecule junctions. Here, we investigate electron transport through an anthraquinone based single-molecule three-terminal device. With the use of an electric-field induced by a gate electrode, the molecule is reduced resulting into a ten-fold increase in the off-resonant differential conductance. Theoretical calculations link the change in differential conductance to a reduction-induced change in conjugation, thereby lifting destructive interference of transport pathways. 
\end{abstract}

\begin{figure}
	\includegraphics[width=0.75\textwidth]{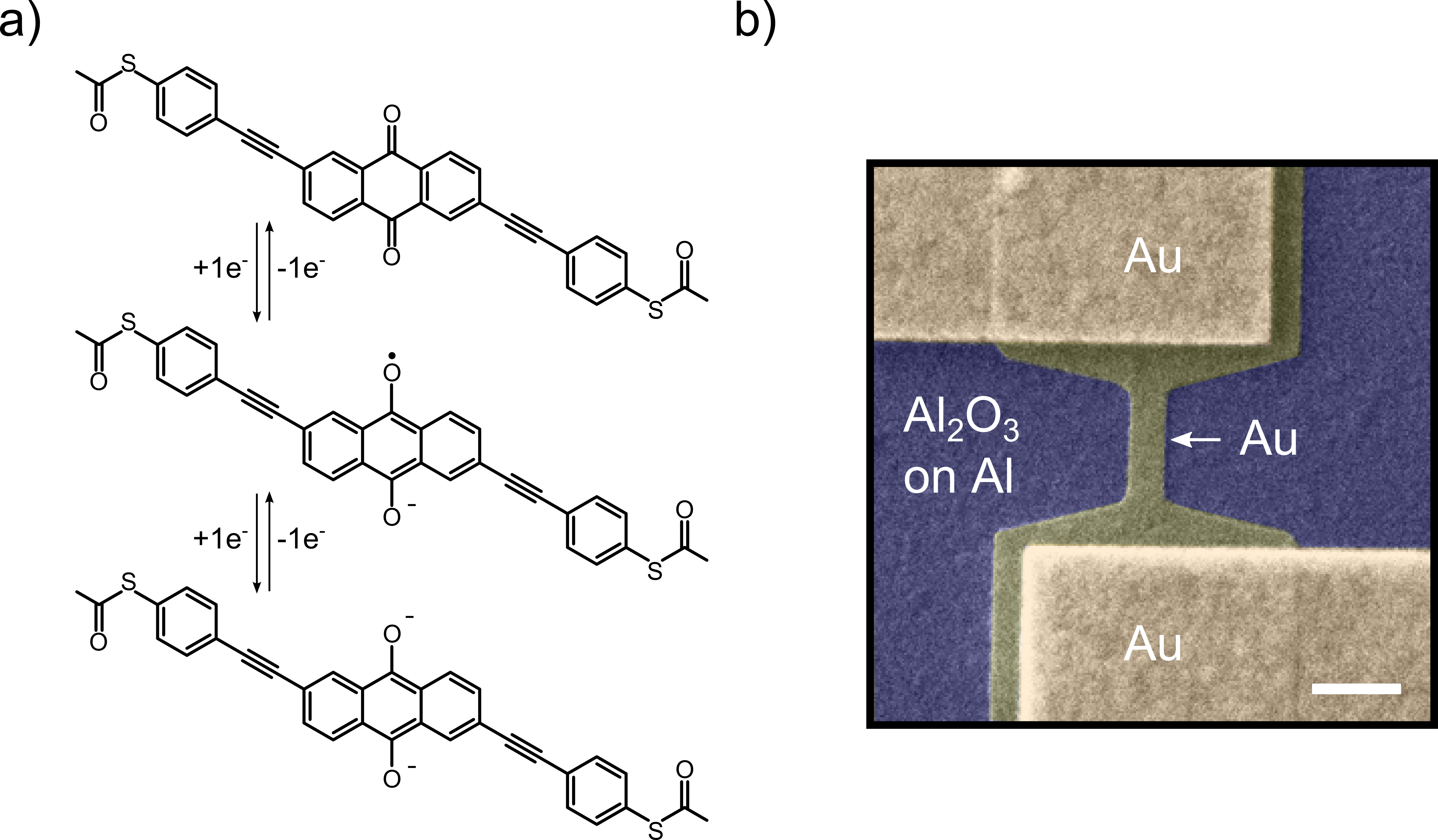}
	\caption{a) Anthraquinone with thiolated spacers. The neutral molecule can be reversibly reduced in two steps, thereby changing the conjugation from cross-conjugation (top) to linear conjugation (bottom). b) Scanning electron microscope image of a break junction on top of an aluminium gate electrode prior to electromigration; scale bar indicates 250 nm.}
	\label{fig:molfab}
\end{figure}

Electric-field control by a gate electrode is a promising route to manipulate charge transport through single molecules\cite{Park2000}. Detailed spectroscopy as a function of bias and gate voltage is possible, enabling the systematic study of electron transport through single-molecule junctions; examples of which are Coulomb blockade\cite{Park2000}, the Kondo effect\cite{Park2002,Liang2002}, vibrational excitations \cite{Yu2004,Osorio2007} and electronic excitations\cite{Osorio2007a}. All these experiments can be essentially understood by considering transport channels through individual orbitals. In some molecules, however, transport may involve more than one orbital at the same time, possibly leading to interfering pathways\cite{Sautet1988,Solomon2008}; changing the occupancy can then induce dramatic changes in the conductance\cite{Pedersen2014}. In quinone-type molecules, this phenomenon is traditionally understood as a change in the conjugation of the pi-electron system (see figure~\plainref{fig:molfab}a). Transport experiments in which the occupancy is changed have been performed at room-temperature in solution using electro-chemical setups \cite{Tsoi2008,Darwish2012a,Baghernejad0}, however these lack the direct electric-field control which can be attained with a gate electrode in solid-state devices. 

We investigate the effect of the electric-field on charge transport through a molecule with an anthraquinone core (AQ). The molecule has spacers terminated by sulfur groups\cite{Dijk2006} for binding to gold electrodes. The neutral molecule is shown in figure~\plainref{fig:molfab}a together with its first two reduced states. It is designed as a redox switch, where the neutral state is in a cross-conjugated form and the twice-reduced state has a linear conjugation. In \citeauthor{Dijk2006}\cite{Dijk2006} it is shown that these two reductions of AQ are reversible and change the electronic structure of the molecule. Furthermore, the occurrence of quantum interference in AQ and its functioning as a redox switch have been investigated theoretically\cite{Markussen2010,Garcia-Suarez2014}. On the experimental side conducting force microscopy\cite{Guedon2012}, eutectic Ga-In top contacting\cite{Fracasso2011} and mechanically controlled break-junctions\cite{Kaliginedi2012} have shown that AQ has a suppressed conductance, which is linked to its cross-conjugation. However, in-situ switching of the conjugation as a function of electric-field has not yet been reported. Here we study AQ in a solid-state three-terminal device and demonstrate electric-field switching of molecular conjugation.

To create three-terminal metal/molecule/metal junctions, gold wires are deposited on top of an oxidized aluminium gate; the result is shown in figure~\plainref{fig:molfab}b. On a silicon chip 22 of such junctions are typically fabricated. A solution of dichloromethane with a concentration of 0.1 mM of AQ is then deposited on the chip. In this solution and at room-temperature, the gold wire is controllably thinned down by electromigration\cite{Strachan2005} until its resistance is about 5 K$\Omega$. Subsequently, the gold wires are left to self-brake\cite{ONeill2007} until the resistance of the junction is between 100 K$\Omega$ and 1 M$\Omega$; this takes one to two hours depending on the thickness of the gold wire. To prepare for cool-down, the sample space is pumped down, evaporating the dichloromethane. The chip is then cooled to 4 K in a liquid helium dewar. Measurements are performed by applying a voltage on the gate electrode and across the source electrodes while measuring the current at the drain. The experimental system possesses a heater-resistor and a 1K pot, which makes temperature-dependent measurements possible. 

\begin{figure}
	\includegraphics[width=1\textwidth]{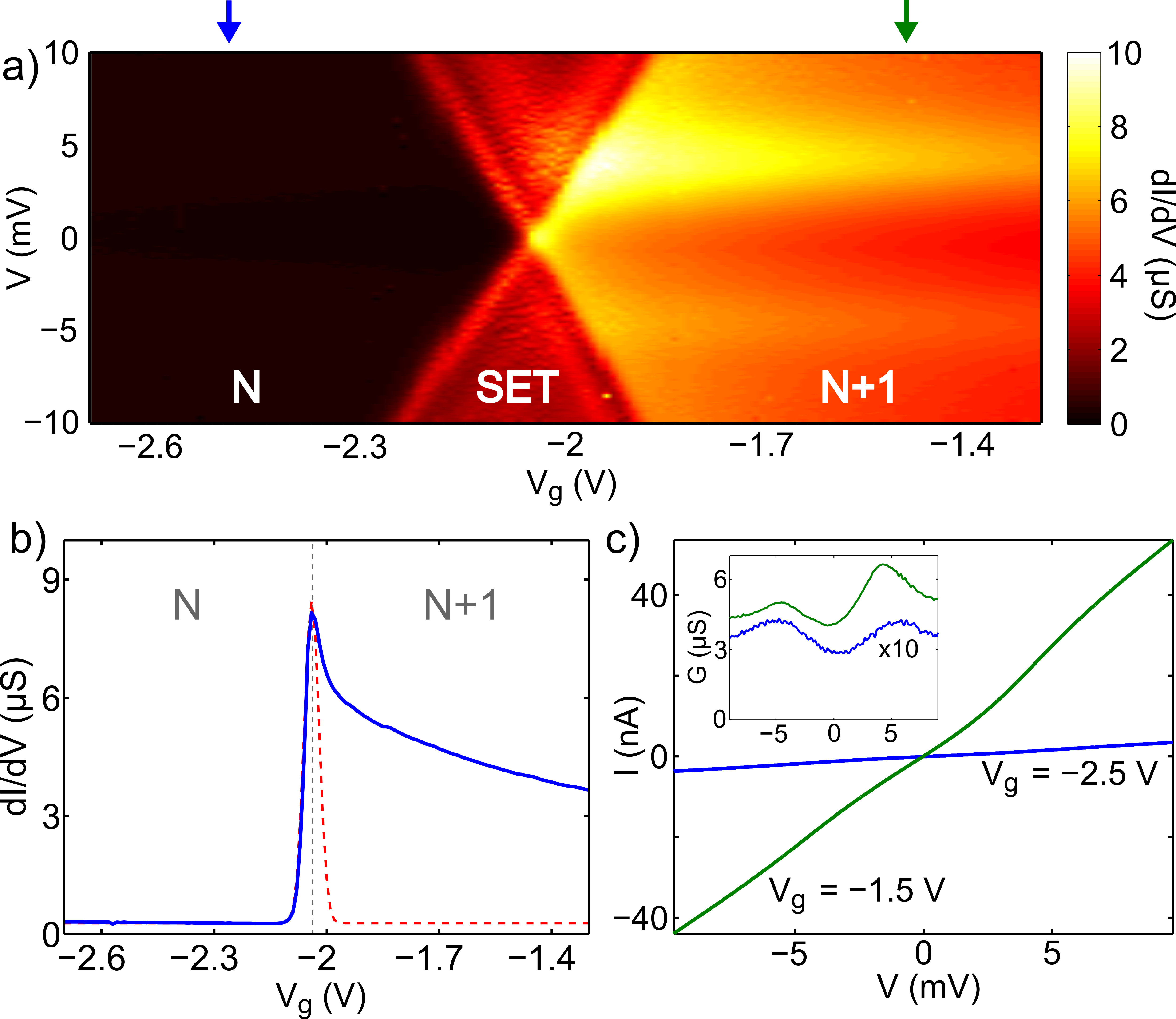}
	\caption{Bias and gate voltage dependent measurement of sample A at 2.9 K. a) Numerical derivative (dI/dV) of the current as a function of bias (V) and gate (V\textsubscript{g}) voltage. Two charge-states are present (labelled N and N+1) and the on-resonant transport regime is indicated by the two red triangles (labelled SET). On the left the off-resonant differential conductance is low, as indicated by the black color. On the right, the red/yellow colors represents a much higher off-resonant differential conductance in the adjacent charge-state. The blue and green arrows indicate the location of the IV traces in figure c). b) Blue line: differential conductance versus gate voltage at V = 0 mV. The red dashed line is a Gaussian function with an electronic coupling, $\Gamma$, of 1.2 meV. c) Current versus bias voltage at the two different gate voltages shown by the arrows in figure a). The inset shows the corresponding differential conductance. The differential conductance of the trace at V\textsubscript{g} = -2.5 V is multiplied by 10.}
	\label{fig:measurements}
\end{figure}

In this paper, we discuss five different junctions that show a pronounced charge-state dependence of the off-resonant differential conductance. Figure~\plainref{fig:measurements}a shows the differential conductance map as a function of bias (V) and gate (V\textsubscript{g}) voltage of one of the five junctions. In this sample (which we will call sample A from now on) two slanted lines cross at V\textsubscript{g} = -2.1 V. The two red-coloured triangles originating from this crossing point indicate areas in which transport is dominated by resonant single electron tunnelling (SET). In the regions left and right of the vertical pair of triangles, transport is off-resonant involving higher-order co-tunnel processes. The regions of off-resonant transport left and right of the resonance differ in charge by one electron on the molecule; as denoted by the transition from N to N+1, where N is the excess charge on the molecule, with the most likely scenario N=0 (see also below). Comparing the differential conductance of the charge-states, it can be clearly seen that the level of differential conductance in the right charge-state is about an order of magnitude higher than that of the left charge-state.

The differential conductance as a function of gate voltage at zero bias is shown in figure~\plainref{fig:measurements}b and clearly shows the difference in off-resonant differential conductance between the two charge-states. At gate voltages below V\textsubscript{g} = -2.1 V the differential conductance is around G = 0.27 $\upmu$S. Increasing the gate voltage above V\textsubscript{g} = -2 V, the differential conductance increases to about G = 2.5 $\upmu$S far from the resonance (see figure S4a in the supplementary information). The peak near V\textsubscript{g} = -2 V is due to resonant transport. It is striking to see that the resonance itself is strongly asymmetric, contrary from what would be expected from usual quantum dot theory\cite{Beenakker1991a,Cuevas2010}. The left side of the resonance follows a Gaussian line-shape (red dashed line in fig~\plainref{fig:measurements}b), but the right side deviates strongly from the Gaussian line-shape. For the other samples an asymmetry is also found, albeit less pronounced (see supplementary information).

Figure~\plainref{fig:measurements}c shows two current-voltage characteristics taken in the left and right charge-state (blue and green arrows in figure~\plainref{fig:measurements}a). The inset shows that at any bias voltage the differential conductance increase is more than an order of magnitude. Figure S1a in the supplementary info shows that the above described features persist even at higher bias voltages up to 40 mV.

The inset of figure~\plainref{fig:measurements}c also shows the presence of excitations of a few meV in the off-resonant transport regime of both charge-states (the one in the N charge-state is not visible with the contrast settings used to obtain figure 2a). This observation of excitations in both off-resonant transport regimes (N and N+1), combined with the fact that excitations with the same energy are present in the resonant transport regime, supports the conclusion that the features in figure~\plainref{fig:measurements} are from the same molecule (see figure S2 in the supporting information). This is further strengthened by figure S1b in the supporting info, which shows the emergence of a weak zero-bias peak in the right charge-state, indicative of Kondo physics associated with an unpaired electron\cite{Goldhaber-Gordon1998,Liang2002,Park2002}. It suggests that by going from the left to the right charge-state, a single electron is added to the system which changes the occupation from even to odd. Since it is unlikely that many electrons can be put on a small molecule, we assign the even occupation to the neutral state (N=0) and the odd occupation to the singly reduced state.

\begin{table}
\centering
\begin{tabular}{l|m{2.5cm}|m{2.5cm}|m{2.5cm}}
\small{\textbf{Sample}} & \small{\textbf{Differential conductance G(N) ($\upmu$S)}} & \small{\textbf{Differential conductance G(N+1) ($\upmu$S)}} & \small{\textbf{Differential conductance difference G(N+1)-G(N) ($\upmu$S)}} \\ 
\hline 
A & 0.27 & 2.54 & 2.27 \\
B & 0.75 & 2.14 & 1.39 \\
C & 0.27 & 2.38 & 2.11 \\ 
D & 0.43 & 2.71 & 2.28 \\ 
E & 6.98 & 8.77 & 1.79
\end{tabular}
\caption{Off-resonant low-bias differential conductance in the N and N+1 charge-state for five different samples. The last column shows the difference in differential conductance between the two charge-states. All values are in $\upmu$S. The off-resonant differential conductance is determined by fitting the left and right side of the N to N+1 resonance with a Lorentzian (Gaussian for sample A) and determining the offset conductance. For Sample A the N+1 differential conductance was estimated by determining the lowest value it reaches in the N+1 state.}
\label{tab:junctions}
\end{table}

Four other samples have also been identified which show a charge-state dependent increase of off-resonant transport when comparing adjacent charge-states. All these samples, including sample A, are shown in figure S4 and S5 in the supplementary information. Table~\plainref{tab:junctions} lists the conductance in the left charge-state and adjacent charge-state to the right (N to N+1 transition) for these five samples. It can be seen that the conductance increase is approximately the same for all junctions.

\begin{figure}
	\includegraphics[width=1\textwidth]{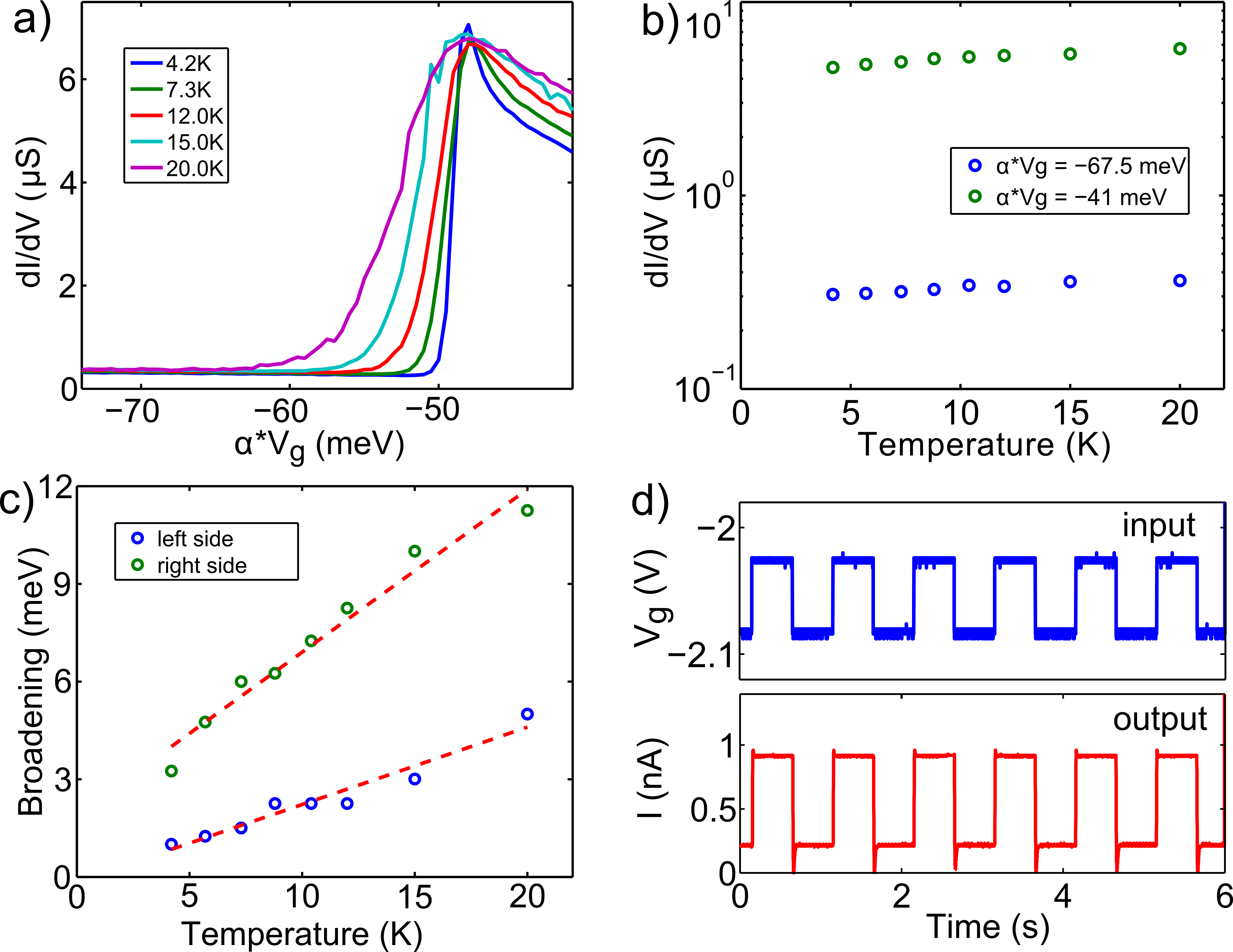}
	\caption{Temperature dependence and switching of sample A. a) Temperature dependence of the differential conductance versus gate at V = 0 mV. Gate is multiplied with $\alpha$ = 25 mV/V, so that the x-axis represents energy scale in meV. b) Log-plot of the differential conductance as a function of temperature at two gate voltages. c) Broadening of the left and right side of the resonance as a function of temperature. Red dashed lines are linear fits to the data with broadening$= c_{1}+ c_{2} *T$ where $c_{1}=-0.006, c_{2}=0.01$ for the left side and $c_{1}=0.08, c_{2}=0.02$ for the right side. d) Conductance switching of the junction at 2.9 K. A square wave is applied on the gate electrode as input. The output is the source-drain current at V = 1 mV.}
	\label{fig:investmeas}
\end{figure}

To further investigate sample A, the temperature dependence of the differential conductance versus gate voltage at V = 0 mV is shown in figure~\plainref{fig:investmeas}a. The figure shows that the resonance widens as the temperature is increased whereas the off-resonant differential conductance remains almost the same. It should be noted that the asymmetric shape persists for all temperatures. Figure~\plainref{fig:investmeas}b compares the off-resonant differential conductance in the two charge-states, showing a slight increase with temperature. This slight increase indicates that Kondo physics is not the main cause of the conduction increase in the right charge-state, as in that case a logarithmic decrease is expected. To quantify the widening of the resonance we have plotted in figure~\plainref{fig:investmeas}c the width of the left side at half maximum and the right side at 3/4th maximum of the resonance. The width on both sides of the resonance has a linear dependence with temperature. Linear dependence is expected when a life-time broadened resonance is temperature broadened by the Fermi function in the leads\cite{Cuevas2010}.

It is interesting to note that the redox-induced conductance switching can be performed reversibly at speeds limited by the RC time of the set-up. Figure~\plainref{fig:investmeas}d shows the result when changing the gate voltage in a square-wave fashion on the time scale of seconds. In this experiment, the square wave signal on the gate was centered around the resonance gate voltage, and as the lower panel shows, it results in a square wave output signal. Figure S7 in the supplementary information shows a 10 Hz sine wave modulating the current output of the junction.

The difference in conductance between the N and N+1 charge-states resembles the transmissions obtained from theoretical calculations on molecules providing two interfering pathways for electrons \cite{Solomon2008,Markussen2010,Valkenier2014}. Quinone structures are examples of such structures\cite{Strange15}. Indeed, these calculations invariably show a significant increase of the conductance upon reduction, even in the off-resonant regime when Coulomb blockade can be significant\cite{Pedersen2014}.

\begin{figure}
	\includegraphics[width=1\textwidth]{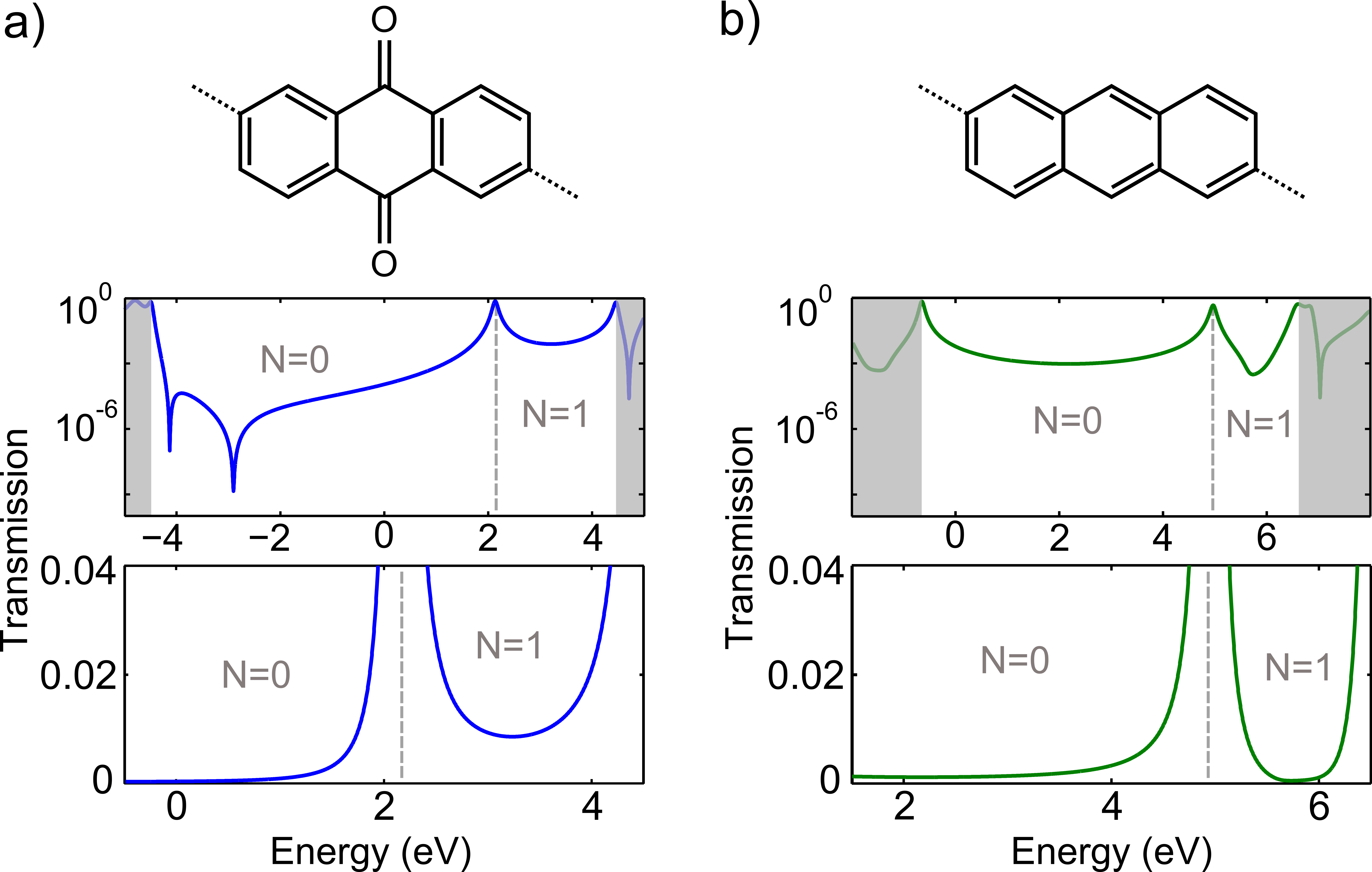}
	\caption{Transmission of the anthraquinone and anthracene group, calculated using the GW method with Ohno parametrization\cite{Ohno64}(parameters taken from \citeauthor{Pedersen13}\cite{Pedersen13} and \citeauthor{roos1967}\cite{roos1967}) a) Transmission for the anthraquinone group. Top figure: Anthraquinone group, dashed lines are the entry and exit site for the transmission calculation. Middle figure: Transmission on a logarithmic scale. The N=0 and N=1 charge-states are identified and the dashed line marks the boundary between the charge-states i.e. it marks the position of the LUMO. the HOMO is visible at an energy of -4.5 eV. Bottom figure: Transmission on linear scale zoomed-in on the reduction (LUMO) peak. b) Transmission of the anthracene group. Top figure: Anthracene group, dashed lines are the entry and exit site for the transmission calculation. Middle figure: Transmission on a logarithmic scale. The N=0 and N=1 charge-states are identified and the dashed line marks the boundary between the charge-states i.e. it marks the position of the LUMO. the HOMO is visible at an energy of -0.7 eV. Bottom figure: Transmission on linear scale zoomed-in on the reduction (LUMO) peak.}
	\label{fig:theory}
\end{figure}

We have performed GW calculations for the pi-system of the central anthraquinone (AQ) and anthracene (AC) group with the Ohno parametrization for the Pariser-Par-Pople (PPP) hamiltonian\cite{Ohno64}. The GW implementation in the GPAW code was used\cite{Enkovaara10,Thygesen08,Strange11}, with the parameters $U=11$~eV and $t=-2.5$~eV for the PPP Hamiltonian and $V=-10$~eV for the oxygen\cite{roos1967,Pedersen13}. Figure~\plainref{fig:theory} shows the calculated transmission through the AQ and AC groups.

The transmission of the AQ group (figure~\plainref{fig:theory}a) clearly shows an increase in the off-resonant conductance upon reduction from the N to N+1 charge-state, both on the linear (lower panel) and logarithmic scale (upper panel). In addition, two interference dips can be identified in the N=0 charge-state indicative of destructive quantum interference\bibnote{In DFT calculations, the lowest (in energy) interference dip is usually found just \emph{below} the HOMO peak\cite{Valkenier2014,Strange15}}. Although it is often difficult to come to a quantitative agreement in the field of molecular electronics, it is interesting to note that the experimental charge-state dependent differential conductance increase of approximately $1\sim 2 \: \upmu$S (table~\plainref{tab:junctions}) is of the same order of magnitude as is present in the calculations (the increase in transmission is approximately T$=0.01$, which corresponds to a conductance increase of $1 \upmu$S).

An important question is whether indeed the quinone-character is responsible for the observed behaviour. To verify this, we also calculated the transmission for anthracene (figure~\plainref{fig:theory}b). This molecule, which lacks the oxygen stub-structures characteristic for quinone, shows a small decrease in the transmission upon reduction and lacks the quantum interference dip structure in the N=0 charge-state. This supports the conclusion that quantum interference is responsible for the charge-state dependent off-resonant differential conductance measured in the anthraquinone molecule.

Comparing the results for the device analysed in this paper and the four presented in the supplementary information shows that, in addition to the common feature of a conductance increase upon reduction, there is device-to-device variation. The explanation for such a variation  can be found in the non-reproducibility of the coupling between molecules and electrodes, and in image charge effects~\cite{Ishii99}. Calculations incorporating these effects are shown in the supporting info (figure S8).

To conclude, we have shown that when changing the electron occupation of an anthraquinone-based single-molecule junction the off-resonant differential conductance changes dramatically. This is accompanied with an asymmetric resonance between the two charge-states. Both effects are linked to an occupation dependent conjugation of anthraquinone; this is supported by GW calculations. This research thus realizes the in-situ change of conjugation in a solid-state single-molecule system and opens up the way to detailed experimental investigation of pi-conjugation based transport phenomena and interfering transport pathways in single molecules.

\begin{acknowledgement}
This research was performed with financial support from the Netherlands Organization for Scientific Research (NWO/OCW), FOM and by an ERC advanced grant (Mols@Mols). We acknowledge M. Kokken for his help with the measurements.
\end{acknowledgement}

Additional measurements on all five samples and theoretical calculations are included in the supporting information. 

\bibliography{anthraquinone}

\end{document}